\documentclass[twocolumn]{aastex62}

\graphicspath{{./}{figures/}}

\received{}
\revised{}
\accepted{}

\submitjournal{AJ}

\shorttitle{Accretion rate and infrared excess in Herbig Ae/Be stars}
\shortauthors{Arun et al.}
\usepackage{subfigure}
\usepackage{mathtools}
\usepackage{multirow}\usepackage[normalem]{ulem}
\useunder{\uline}{\ul}{}

\begin{document}

\title{On the mass accretion rate and infrared excess in Herbig Ae/Be Stars}

\correspondingauthor{Blesson Mathew}
\email{blesson.mathew@christuniversity.in}

\author{R. Arun}
\affil{Department of Physics and Electronics, CHRIST (Deemed to be University), Bangalore 560029, India}
\author{Blesson Mathew}
\affil{Department of Physics and Electronics, CHRIST (Deemed to be University), Bangalore 560029, India}
\author{P. Manoj}
\affiliation{Department of Astronomy and Astrophysics, Tata Institute  of Fundamental Research, Homi Bhabha Road, Colaba, Mumbai 400005, India}
\author{K. Ujjwal}
\affil{Department of Physics and Electronics, CHRIST (Deemed to be University), Bangalore 560029, India}
\author{Sreeja S. Kartha}
\affil{Department of Physics and Electronics, CHRIST (Deemed to be University), Bangalore 560029, India}
\author{Gayathri Viswanath}
\affil{Department of Physics and Electronics, CHRIST (Deemed to be University), Bangalore 560029, India}
\author{Mayank Narang}
\affiliation{Department of Astronomy and Astrophysics, Tata Institute of Fundamental Research, Homi Bhabha Road, Colaba, Mumbai 400005, India}
\author{K.T. Paul}
\affil{Department of Physics and Electronics, CHRIST (Deemed to be University), Bangalore 560029, India}

\begin{abstract}

The present study makes use of the unprecedented capability of the {\it Gaia} mission to obtain the stellar parameters such as distance, age, and mass of HAeBe stars. The accuracy of {\it Gaia} DR2 astrometry is demonstrated from the comparison of the {\it Gaia} DR2 distances of 131 HAeBe stars with the previously estimated values from the literature. This is one of the initial studies to estimate the age and mass of a confirmed sample of HAeBe stars using both the photometry and distance from the {\it Gaia} mission. Mass accretion rates are calculated from $H\alpha$ line flux measurements of 106 HAeBe stars. Since we used distances and the stellar masses derived from the {\it Gaia} DR2 data in the calculation of mass accretion rate, our estimates are more accurate than previous studies. The mass accretion rate is found to decay exponentially with age, from which we estimated a disk dissipation timescale of $1.9\pm 0.1$ Myr. Mass accretion rate and stellar mass exhibits a power law relation of the form, $\dot{M}_{acc}$ $\propto$ $M_{*}^{2.8\pm0.2}$. From the distinct distribution in the values of the infrared spectral index, $n_{2-4.6}$, we suggest the possibility of difference in the disk structure between Herbig Be and Herbig Ae stars. 

\end{abstract}


\section{Introduction} 
\label{sec:intro}

Herbig Ae/Be stars are intermediate-mass pre-main sequence (PMS) stars with masses between 2 and 10 M\textsubscript{\(\odot\)}. They are often used to understand the missing link in the star formation sequence connecting T Tauri stars and massive young stellar objects (e.g. \citealp{Herbig1960, WATERS1998, Oudmaijer2017}). Herbig Ae/Be stars (hereafter HAeBe) show emission lines in their spectrum and exhibit infrared excess (known as IR excess) in the continuum, suggestive of hot and/or cool dust in the circumstellar medium (CSM) \citep{Hillenbrand1992, Malfait1998}. The emission lines such as H$\alpha$ are formed in the CSM and are used for understanding the mass accretion process in HAeBe stars (eg. \citealp{Hamann1992, VIEIRA2003, manoj2006, mendi2011, mENDI2011A}). 

Understanding the accretion of material from the CSM is important to study the PMS evolution because it can provide vital information about the formation and evolution of planets around the stars \citep{Muzerolle2003, Beltran2016}. It is proposed that Herbig Ae (HAe) and Herbig Be (HBe) stars may show considerable differences in disc morphology and mode of accretion \citep{Vink2002, Alonso2009, Vioque2018}. However, in order to establish these results, we need to have precise distance measurements. This is due to the fact that the precision of stellar parameters such as age, mass, log(g) etc., strongly depend on precise distance measurements. One of the pioneering missions which provided accurate distances of nearby astronomical objects was the Hipparcos mission. Based on the distance measurements of nearby HAeBe stars from the Hipparcos mission \citep{ESA1997}, \cite{VANDEN1998} derived the astrophysical parameters of a sample of 44 HAeBe stars and found that 65\% of HAeBe stars show photometric variability. It may be noted that Hipparcos provided reliable distance values for stars within 1 kpc to the Sun \citep{deZEEuw1999}. The {\it Gaia} mission is designed to provide high-quality astrometry and photometry of 1.3 billion stars \citep{gaia2016, GaiaColla2016}. With the second data release of {\it Gaia} (named as {\it Gaia} DR2) \citep{gAIANEW2018}, it is possible to get parallax measurements of stars with uncertainties limited to 0.04 mas, for sources brighter than G = 14 mag \citep{Luri2018}. From precise distance measurements, it is possible to derive the relations connecting the IR excess and mass accretion rates ($\dot{M}_{acc}$) with the stellar parameters of HAeBe stars. This can be used to understand whether magnetospheric or disc accretion plays a major role in HAeBe stars.   

 In this work, we estimate the stellar parameters of a well-studied sample of HAeBe stars, thereby understanding the mass accretion process in pre-main sequence stars.  We present the sample of HAeBe stars used for this study in Sect. 2. The results of this study are presented in Sect. 3, wherein we discuss the procedure associated with distance and extinction measurements. Also, we estimate the mass and age of HAeBe stars and discuss mass accretion in HAeBe stars. Recently, \cite{Vioque2018} estimated stellar parameters of HAeBe stars using distance measurements from  {\it Gaia} DR2. They based their analysis on the derived quantities such as luminosity and temperature, which can introduce additional errors in the estimation of mass and age of HAeBe stars. Instead, in the present study, we based the analysis on {\it Gaia} color-magnitude diagram. The main results are summarized in Sect. 4.  

\section{Data Inventory}
\label{sec:data}

A sample of 142 stars is taken from  \cite{Mathew2018}, which is a carefully selected, well-studied sample of HAeBe stars from \cite{The1994}, \cite{manoj2006}  and \cite{FAirlamb2015}. \cite{Mathew2018} discussed various mechanisms for the formation of O{\sc i} emission lines in HAeBe stars and found that Lyman beta fluorescence is the dominant excitation mechanism. This is the second work in the series, studying about the $\dot{M}_{acc}$ and IR excess in HAeBe stars. Here we re-estimate the relations connecting the $\dot{M}_{acc}$ with the stellar parameters such as age and mass in the context of the {\it Gaia} DR2 release. These new estimates will be used for our future work to explore the possibility of using O{\sc i} 8446 \AA~emission line as an accretion indicator in HAeBe stars (Mathew et al. in prep.).   

The coordinates, proper motions and $V$ magnitudes of the 142 stars are taken from the literature. RA and Dec of these stars are converted from J2000 to J2015.5 epoch using their proper motion. A query for a {\it Gaia} DR2 match for these stars was then performed around the converted coordinates with a search radius of 10 arcsec via the Mikulski Archive for Space Telescopes (MAST)\footnote{https://archive.stsci.edu/}. If a match was not found, then the search radius was increased up to 30 arcsec. This procedure returned 354 {\it Gaia} DR2 rows for 142 stars. For 60 stars, only one {\it Gaia} DR2 match was returned. For the remaining 82 stars with multiple entries, those which had $\mid$G$-$V$\mid$ mag  $>$ 3.5 were removed. For the remaining multiple entries, the {\it Gaia} DR2 row with the closest positional match was selected for which $\mid$G$-$V$\mid$ mag $\leq$ 2. Thus we got the {\it Gaia} DR2 parallax and magnitudes for all stars in the sample. After avoiding 11 sources, where 6 showed no parallax data and 5 had negative parallax, we finalized our sample of HAeBe stars to 131. These stars are found in the distance range of 0.09$-$6 kpc, with a range in {\it Gaia} G band magnitude from 4.4 to 14.5 mag. 
\section{Results}

\subsection{Comparison of the {\it Gaia} DR2 distances with previous estimates}

The uncertainty in the distance determination of stars is mitigated to a considerable extent due to the precision of the {\it Gaia} mission. Although {\it Gaia} DR2 provides accurate positions and parallax measurements via a rigorous astrometric reduction technique, the estimation of distance by simple inversion of {\it Gaia} parallax does entail certain inherent problems. The distance obtained through such a method is acceptable only when the parallax measurements are fairly precise, i.e., when the signal to noise ratio (SNR) of the parallax measurement is preferably high (SNR$\geq$5). In cases where fractional parallax uncertainty is high, the probability distribution for the distance inferred from inverted parallax becomes strongly asymmetric and non-Gaussian in nature. Furthermore, the distance thus estimated will be nonphysical if the concerned parallax measurement is negative, owing to the large measurement noise or due to the star moving opposite to the direction of the true parallactic motion. To tackle this problem, \cite{bailerjohns2018} applied a probabilistic approach to estimate distances to 1.3 billion stars having {\it Gaia} DR2 data. They adopted the distance likelihood (inferred from {\it Gaia} parallax) and a distance prior (an exponentially decreasing space density prior that is based on a Galaxy model) approach. The distance estimates and corresponding uncertainties thus determined are purely geometric and devoid of any underlying assumptions. Hence, for the present study, we use the distance estimates from \cite{bailerjohns2018}, which are listed in \autoref{tab:Table1}.  

\begin{figure}[htp]

    \includegraphics[width=1\columnwidth]{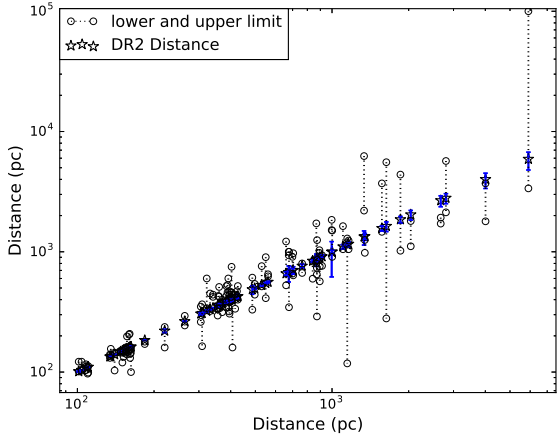}
 
    \caption{Figure shows the comparison between the distances of HAeBe stars from the {\it Gaia} DR2 with the values from previous studies. The distance of the HAeBe stars in parsec is shown in both the axes. Distance estimated from the {\it Gaia} DR2 parallax,  using the method outlined in Sect. 3.1, is shown in star symbol, with the error shown in blue line. The lower and upper bound values of distance for each star is compiled from the literature and is shown as two open circles connected by a dotted line.}
    \label{fig:1}
\end{figure}

We compared the distance estimated from the {\it Gaia} DR2 with the values listed in the literature. \cite{manoj2006} compiled the distances of HAeBe stars from various studies and provided the best estimate of distance for each star. This is supplemented with the distance information from the {\it Gaia} DR1 \citep{GaiaColla2016} and those given in \cite{FAirlamb2015}. The extreme values of distance from these compilations are included in \autoref{fig:1} along with the {\it Gaia} DR2 estimates. It can be seen from the figure that distance estimate from the {\it Gaia} DR2 is more accurate (with minimal error) than previous estimates.  

\subsection{Extinction Calculation}

The extinction in all the photometric bands, G, G\textsubscript{BP} and G\textsubscript{RP}, are listed in the {\it Gaia} archive. But this extinction and reddening values are limited to a small number of objects. The extinction calculation is done by an automated algorithm, which is explained in detail in \citet{Evans2018}. Also, they have listed the caveats involved in the automated way of estimating extinction values. For this work, we have independently estimated the extinction values from the extinction curve of \cite{Mclure2009}. From the curve we calculated $\bigg[\displaystyle\frac{\mbox{$A_G$}}{\mbox{$A_V$}}\bigg]$,$\bigg[\displaystyle\frac{\mbox{$A_{G_{BP}}$}}{\mbox{$A_V$}}\bigg]$
 and $\bigg[\displaystyle\frac{\mbox{$A_{G_{RP}}$}}{\mbox{$A_V$}}\bigg]$.

The A\textsubscript{V} values for our sample of HAeBe stars are taken from \cite{FAirlamb2015}, \cite{2016NewA...44....1C} and \cite{Mathew2018}. \cite{2004AJ....127.1682H} suggested using high values of total-to-selective extinction (R\textsubscript{V} = 5) for estimating the extinction values of HAeBe stars. This is suggestive of grain growth in the disk of HAeBe stars \citep{Gorti1993,manoj2006}. For the present work, we adopted R\textsubscript{V} = 5 while calculating the extinction (A\textsubscript{V}) values. This method was followed while calculating the A\textsubscript{V} values of HAeBe stars in \cite{Mathew2018}. Hence, for this analysis, we included the A\textsubscript{V} values of HAeBe stars which are listed in \cite{Mathew2018}. For remaining stars, A\textsubscript{V} values are taken from \cite{FAirlamb2015} and \cite{2016NewA...44....1C}, which are re-estimated for R\textsubscript{V} = 5. It may be noted that \cite{2004AJ....127.1682H} pointed out that the age and luminosity of HAeBe stars better match with that of PMS stars when R\textsubscript{V} = 5 is employed. The  A\textsubscript{V} values estimated for all the HAeBe stars will be used for correcting the {\it Gaia} photometry for extinction. 

The mean wavelength values in the {\it Gaia} passbands and Johnson $V$ band are taken from \cite{jordi2010}. The $\bigg[\displaystyle\frac{\mbox{$A_G$}}{\mbox{$A_V$}}\bigg]$,$\bigg[\displaystyle\frac{\mbox{$A_{G_{BP}}$}}{\mbox{$A_V$}}\bigg]$
 and $\bigg[\displaystyle\frac{\mbox{$A_{G_{RP}}$}}{\mbox{$A_V$}}\bigg]$ values for different ranges of A\textsubscript{V} are calculated using \cite{Mclure2009}, which are listed below.

\vspace{0.5cm} 
For $A_V$ $\leq$ 2.5
\begin{equation}
\frac{A_G}{A_V} = 0.831, \quad
\frac{A_{G_{BP}}}{A_V} = 1.032, \quad
\frac{A_{G_{RP}}}{A_V} = 0.678
\end{equation} 

For 2.5 $<$ $A_V$ $<$ 7.5
 
\begin{equation}
\frac{A_G}{A_V} = 0.831, \quad
\frac{A_{G_{BP}}}{A_V} = 1.028, \quad
\frac{A_{G_{RP}}}{A_V} = 0.672
\end{equation} 

For 7.5 $\leq$ $A_V$ 
 
\begin{equation}
\frac{A_G}{A_V} = 0.831, \quad
\frac{A_{G_{BP}}}{A_V} = 1.028, \quad
\frac{A_{G_{RP}}}{A_V} = 0.672,
\end{equation}

Using these relations we estimated $A_G$, $A_{G_{BP}}$ and $A_{G_{RP}}$ from the known values of A\textsubscript{V}. This is further used to correct the {\it Gaia} magnitudes, which will be used for this work.

\subsection{Age and mass of HAeBe stars }

In addition to precise astrometric measurements, the {\it Gaia} DR2 lists three broad-band photometric magnitudes, G, G\textsubscript{BP} and G\textsubscript{RP}, extinction in G band (A\textsubscript{G}) and reddening (E(G\textsubscript{BP} $-$ G\textsubscript{RP})) values. This provides the possibility to construct a color-magnitude diagram (CMD) exclusively from {\it Gaia} magnitudes \citep{GaiaHR2018}. We identified that the G-band filter in {\it Gaia} is very wide (720 $nm$) and hence can introduce uncertainty in G magnitude measurements. Hence for the present work, we use G\textsubscript{BP} and G\textsubscript{RP} magnitudes for constructing the CMD. The observed {\it Gaia}  G\textsubscript{BP} and G\textsubscript{RP} are corrected for extinction using the method discussed in Sect. 3.2. Further, making use of the distance estimates (see \autoref{tab:Table1}), we estimated the absolute G\textsubscript{RP} magnitude (M$_{G_{RP}}$), which will be used for the CMD analysis. Usually, the construction of the CMD with non-homogeneous datasets belonging to different epochs can introduce systematic errors in the estimation of stellar parameters. The use of {\it Gaia} astrometry and photometry for the CMD analysis alleviate this issue. Also, we derived the age and mass of HAeBe stars from the observed CMD rather than from a theoretical Hertzsprung-Russell (HR) diagram. Luminosity calculation for stars in the HR diagram involves the conversion of V magnitude to luminosity using bolometric corrections. Such a conversion will provide substantial errors in mass and age estimates. In addition, the effective temperature of the star (T\textsubscript{eff}) is identified using a calibration table which introduces degeneracy in T\textsubscript{eff} for relatively nearer spectral types.
\begin{figure}
\includegraphics[width=1.1\columnwidth]{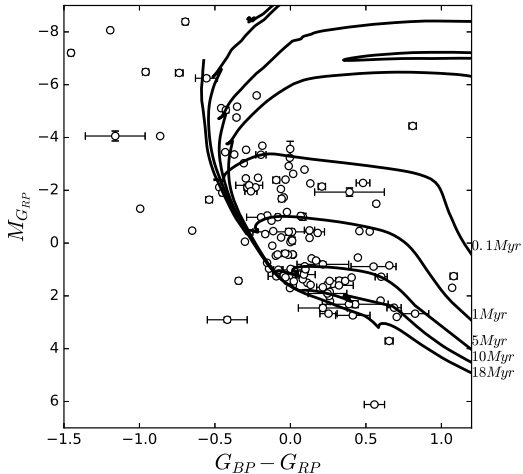}
    \caption{Figure shows MIST isochrones over-plotted on the {\it Gaia} CMD containing 131 HAeBe stars. Isochrones of ages from 0.1 to 18 Myr are plotted in the CMD with metallicity, Z\textsubscript{\(\odot\)} = 0.0152 and (V/V\textsubscript{crit}) = 0.4.}
    \label{fig:2}
\end{figure}
\begin{figure}
\begin{flushleft}
\includegraphics[width=1.1\columnwidth]{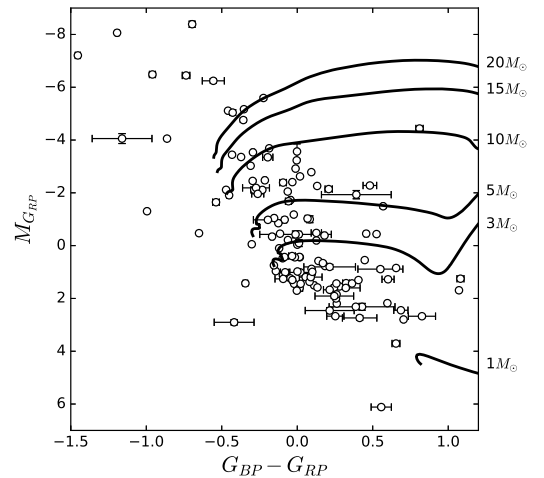}
    \caption{Figure shows the {\it Gaia} CMD containing 131 HAeBe stars over-plotted with the MIST evolutionary tracks. Evolutionary tracks with masses ranging from 1 to 20 M\textsubscript{\(\odot\)} are plotted in the CMD. We used the MIST tracks with metallicity, Z\textsubscript{\(\odot\)} = 0.0152 and (V/V\textsubscript{crit}) = 0.4.}
    \label{fig:3}
    \end{flushleft}
\end{figure}

The age and mass of the HAeBe stars are estimated by plotting the Modules for Experiments in Stellar Astrophysics (MESA) isochrones and evolutionary tracks (\href{http://waps.cfa.harvard.edu/MIST/}{MIST})\footnote{http://waps.cfa.harvard.edu/MIST} \citep{choi2016, Dotter2016} in the {\it Gaia} CMD. The MIST is an initiative supported by NSF, NASA and Packard Foundation which builds stellar evolutionary models with different ages, masses, and metallicities. The updated models in the MIST archive included isochrones and evolutionary tracks for the {\it Gaia} DR2 data. We know that HAeBe stars have a range of rotation rates but we adopted the isochrones corresponding to (V/V\textsubscript{crit}) = 0.4, since that is the only model available in the MIST database for a rotating system. Also, we adopted the metallicity  $\bigg[\displaystyle\frac{\mbox{Fe}}{\mbox{H}}\bigg]$ = 0 (corresponding to solar metallicity; Z\textsubscript{\(\odot\)} = 0.0152) for estimating the age and mass of HAeBe stars.

The {\it Gaia} CMD for our sample of 131 HAeBe stars is shown in \autoref{fig:2} \& \autoref{fig:3}.  From \autoref{fig:2}, we estimated the ages of 110 HAeBe stars by over-plotting MIST isochrones. They are found to be in the range of 0.1 to 15 Myr. From \autoref{fig:3}, it can be seen that the mass range of our sample of HAeBe stars is 1.4 to 25  M\textsubscript{\(\odot\)}. The masses are identified from the coincidence of the data points with the grid of MIST evolutionary tracks. The estimated ages and masses of the HAeBe stars from this work are compared with that in \cite{Vioque2018} and are listed in \autoref{tab:Table1}. We found that 21 stars from our sample are placed below the main sequence and hence the parameters could not be estimated. Since these stars are catalogued as HAeBe stars, they may be properly positioned in the pre-main sequence location in previous studies. HAeBe stars are known to show photometric variability \citep{VANDEN1998}. The stars which are found below the main sequence in \autoref{fig:2} \& \autoref{fig:3} may show photometric variability. Also, some stars are positioned in the evolved region of the evolutionary track. Further studies are needed to evaluate the nature of these candidates.

\subsection{Mass accretion rates of HAeBe stars}

The mass accretion process during the pre-main sequence phase represents one of the important mechanisms associated with star formation. In T Tauri stars, mass accretion is through a process known as magnetospheric accretion (MA) in which the magnetosphere of the host star truncates the circumstellar disk at a few stellar radii and the material from the disk fall on to the star at free-fall velocities along the magnetic field lines, which in turn create shocks at the surface of the star. The hot (10\textsuperscript{4} K) emission from the post-shock gas appear as excess in the UV continuum of T Tauri stars (e.g. \citealp{CalvetandGul1998,Gullbring1998,Hartmann1998,Bouvier2007}). The MA accretion model may not be a viable mode of accretion in HAeBe stars since there are no convincing signatures of a magnetic field in these systems \citep{Alecian2013}. Although many studies suggest disk accretion as the possible mechanism in Herbig Be stars, a consensus is yet to be obtained whether MA accretion can account for mass accretion in low mass HAeBe stars \citep{Muxerolle2004}. For the present work, we employed magnetospheric accretion formalism while calculating the $\dot{M}_{acc}$ in HAeBe stars.
\begin{figure*}[t!]
 \centering   
\subfigure[]{\label{fig:a}\includegraphics[width=1\columnwidth]{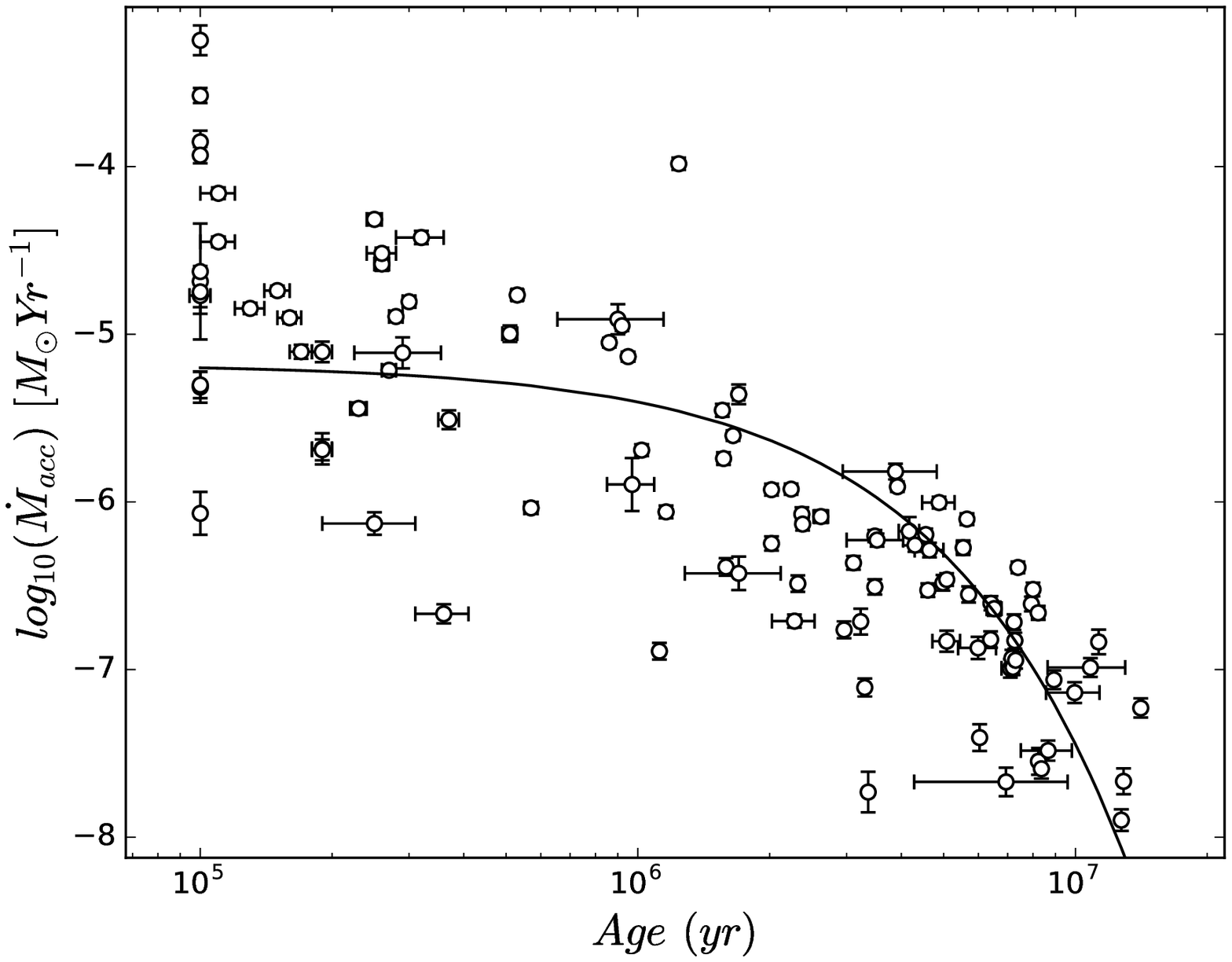}}
\subfigure[]{\label{fig:b}\includegraphics[width=1\columnwidth]{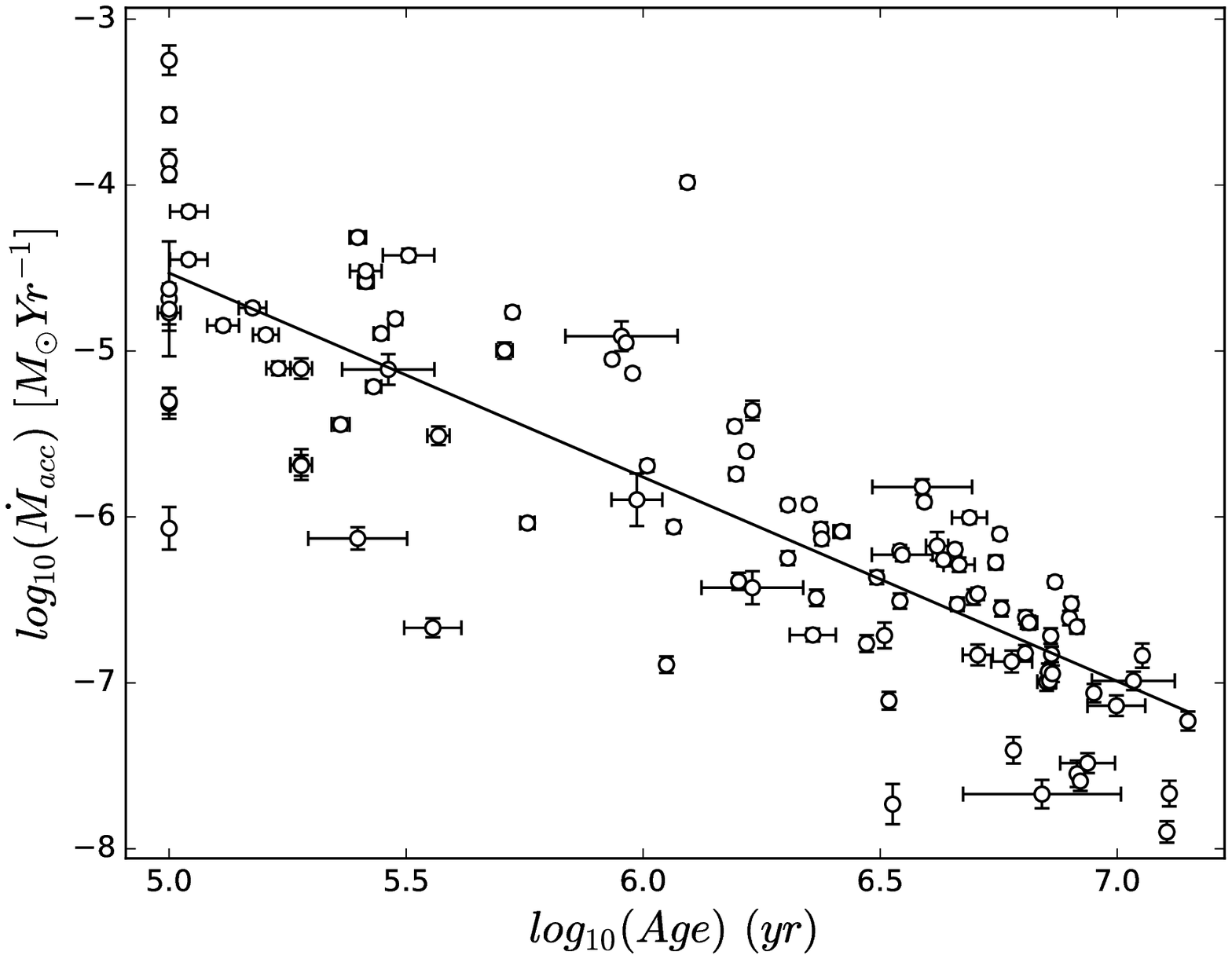}}
\caption{Figures show the relationship between $log(\dot{M}_{acc})$ and age (t) in yr for a sample of 106 HAeBe stars. Figure (a) shows an exponential decrease in $log(\dot{M}_{acc})$  with age for HAeBe stars, as expressed in Eqn. 8. The best fit gives $\tau$ = $1.9 \pm 0.1$ Myr illustrated with a solid line. Figure (b) depicts a log-log plot of $\dot{M}_{acc}$ with age. The power law relation given by Eqn. 9 gives a value of $\eta$ = $1.2\pm 0.1$, which is represented by a solid line.} 
\end{figure*}

The H$\alpha$ line flux values of 102 HAeBe stars are taken from \cite{Mathew2018}, \cite{Fairlamb2017} and \cite{mENDI2011A}. In addition, we took the H$\alpha$ equivalent width (EW) for four stars from \cite{Boehm1995}, \citet{Baines2006}, \citet{BorgesFernandes2007} and \citet{Vieria2011}. The EW is converted to line flux from the $R$ band magnitude using the method mentioned in \cite{Mathew2018}. Hence, for the present analysis, we will be using the H$\alpha$ line flux (F$_{H\mathrm{\alpha}}$) values of 106 HAeBe stars. The H$\alpha$ line flux is converted to luminosity (L$_{H\mathrm{\alpha}}$) using the equation,
\begin{equation}
   L_{H\mathrm{\alpha}}=4\pi d^2 F_{H\mathrm{\alpha}}
\end{equation}

where $d$ is the distance in pc. The accretion luminosity (L\textsubscript{acc}) is calculated using the empirical relation given in \cite{Fairlamb2017}, which is reproduced below. 

\begin{equation}
   log\frac{L_{acc}}{L_{\odot}} = 2.09(\;\pm\;0.06) + 1.00(\; \pm \;0.05) \times log\frac{L_{H\mathrm{\alpha}}}{L_{\odot}}
\end{equation}

The ($\dot{M}_{acc}$) can be derived from the L\textsubscript{acc} using the relation,  

\begin{equation}
L_{acc} = \frac{GM_*\dot{M}_{acc}}{R_*}(1-\frac{R_*}{R_i})
\end{equation}

where $M_*$ is the mass of HAeBe stars, estimated in Sect. 3.3 and given in \autoref{tab:Table1}; $R_i$ is the disk truncation radius. For T Tauri stars, $R_i$ is assumed to be 5 R$_*$ \citep{Gullbring1998,Costigan2014}. HAeBe stars are fast rotators and therefore have a smaller co-rotation radius. The disk truncation radius, $R_i$, should be smaller than the co-rotation radius \citep{Shu199}. Thus in this work, we adopt disk truncation radius, $R_i$ = 2.5 R$_*$ \citep{Muxerolle2004,mendi2011,FAirlamb2015}. The stellar radius $R_*$ for the 106 HAeBe stars are calculated using the equation,

\begin{equation}
R_* = \bigg(\frac{L_*}{4 \pi \sigma T_{eff}^4}\bigg)^{1/2}
\end{equation}

where $L_*$ is the bolometric luminosity of the star, which is calculated from the $V$ magnitude, bolometric correction and {\it Gaia} distance. Using the calibration table listed in \cite{mamajeck2013}, we identified T\textsubscript{eff} and bolometric correction corresponding to the spectral type of the HAeBe star. The $V$ magnitudes of 101 HAeBe stars are compiled from AAVSO Photometric All Sky Survey \citep[APASS;][]{APASS2016} and Tycho-2 \citep{Hogg2000} catalogues. The remaining 5 stars which had no $V$ magnitude listed in both the catalogues are taken from the following references $-$ \citet{Herbst1999}, \citet{Getman2008}, \citet{Fresneau2009} and \citet{Girard2011}.

\subsection{Correlation analysis of mass accretion rates with stellar parameters}

The relationship between the $\dot{M}_{acc}$ and the stellar parameters such as age and mass are analyzed in some of the studies (e.g. \citealp{mendi2011, Mendigutia2015, Fairlamb2017}). However, in the context of precise mass and age estimates using {\it Gaia} DR2, we re-assessed the relations between $\dot{M}_{acc}$ and the stellar parameters using the largest sample of 106 HAeBe stars to date. Figure \autoref{fig:a} illustrates the correlation between the $log(\dot{M}_{acc})$ and age of HAeBe stars. It can be seen that $log(\dot{M}_{acc})$ decays exponentially with the age of HAeBe stars. This trend is discussed in the studies of \cite{manoj2006} and \cite{Mendi2012}. From the rate of decline of accretion rate, it is possible to estimate the disk dissipation timescale,  $\tau$, using the relation,
\begin{equation}
\dot{M}_{acc}(t) = \dot{M}_{acc}(0)e^{-t/\tau} 
\end{equation}

where $t$ is the age of HAeBe stars. By fitting the relation to the set of data points, we obtained the disc dissipation time scale, $\tau$ = $1.9\pm0.1$ Myr. This value is near to that given in \cite{Mendi2012}, which is $\tau$ = $1.3^{+1.0}_{-0.5}$ Myr. It may be noted that $\tau$ for T Tauri stars is 2$-$4 Myr \citep{Fedele2010, Takagi2014}. We find a lower $\tau$ value for HAeBe stars indicating that the disk dissipation timescale is shorter for intermediate mass young stars compared to their lower mass counterparts.

Further, another parameter used in the literature for calculating the rate of decline of accretion rate with age in young stellar objects (YSOs) is the power law index, $\eta$ \citep{Hartmann1998, Mendi2012, FAirlamb2015}. The relation which connects $\dot{M}_{acc}$ with age of the star can also be considered as a power law distribution of the form, 
\begin{equation}
    \dot{M}_{acc} = constant \times t^{-\eta}. 
\end{equation}

From the best fit to the distribution of the data points in Figure \autoref{fig:b}, we obtained $\eta$ = $1.2 \pm 0.1$. This value is on the lower end when compared to the estimates of \cite{Mendi2012} and \cite{FAirlamb2015}, which are $1.8^{+1.3}_{-0.7}$ and $1.92 \pm 0.09$, respectively.  This could be because of the increased number of high mass HBe stars in our sample.  

\begin{figure}[htp]

    \includegraphics[width=1\columnwidth]{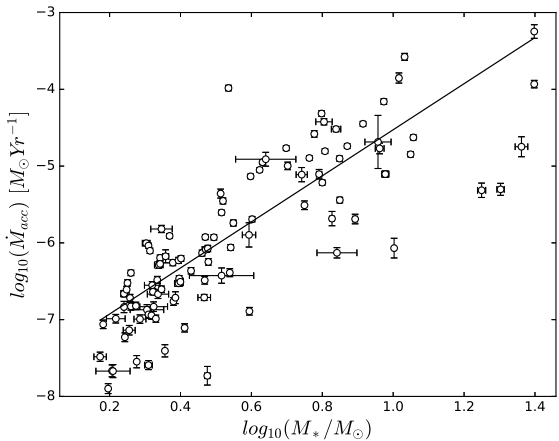}
    \caption{Figure illustrates the log-log plot of $\dot{M}_{acc}$ and stellar mass for a sample of 106 HAeBe stars. The solid line shows the best fit for the power law relation between $\dot{M}_{acc}$ and stellar mass in HAeBe stars. The power law index estimated from the best fit is 2.8$\pm$0.2}
    \label{fig:5}
\end{figure}

In \autoref{fig:5} we plotted the correlation between $\dot{M}_{acc}$ and stellar mass. Our sample of HAeBe stars cover a broader range in spectral type/mass and $\dot{M}_{acc}$ ($\sim$ $10^{-3}$ $-$ $10^{-7}$ $M_{\odot}$ yr$^{-1}$), when compared to the sample of stars given in \cite{mendi2011}. This is because our sample contains high mass candidates with mass $>$~6 $M_{\odot}$, whereas those listed in \cite{mendi2011} are with mass $<$~6 $M_{\odot}$. The best fit for our sample of HAeBe stars in \autoref{fig:5} provides the relation $\dot{M}_{acc}$ $\propto$ $M_{*}^{2.8{\pm0.2}}$. \cite{mendi2011} did a similar study and obtained a steep power law relation, $\dot{M}_{acc}$ $\propto$ $M_{*}^{5}$. The reason for a steeper power law  relation might be due to the unavailability of massive HAeBe stars in their sample. The Pearson correlation coefficient for our fit is 0.81 for a sample size of 106 stars. Incidentally, \cite{FAirlamb2015} obtained the relation between stellar mass and accretion rate as $\dot{M}_{acc}$ $\propto$ $M_{*}^{3.72 \pm 0.27}$, which comes close to our estimate. It may be noted that the mass dependence of accretion rate in T Tauri stars is lower than the value calculated for HAeBe stars, i.e., $\dot{M}_{acc}$ $\propto$ $M_{*}^{2}$ \citep{Muz2005,Natta2006}. 

The best fit and the confidence limits for Figures 4(a), 4(b) and 5 are determined using the Monte Carlo method to account for the associated uncertainties in age, mass and $\dot{M}_{acc}$. For this purpose, 100,000 samples for age, mass and $\dot{M}_{acc}$ were created. The values for these samples were randomly drawn from a Gaussian distribution having a mean equal to the actual measured value in each case and a standard deviation equal to the associated uncertainty. The best fit is then estimated for each of the resulting data set. The fit parameters obtained for all 100,000 datasets results in a normal distribution, the mean of which, along with its 3 $\sigma$ confidence limits, is taken as the final best fit. 

\subsection{Quantifying IR excess using spectral index}

IR excess in the Spectral Energy Distribution (SED) is one of the important criterion used in identifying YSOs.  It provides a better understanding of the composition of gas and dust in the disk of a PMS star. \cite{Lada_Wilk1984} differentiated YSOs into different classes from the shape of their SEDs in the IR region. \cite{LAda1987} quantified the classification scheme using the slope in the IR region of the SED, which are known as Lada indices. The YSOs can be classified as Class 0, Class I, Class II and Class III, based on the steepness of the indices at various wavelength intervals \citep{LAda1987, Andre1993}. The estimation and analysis of Lada indices are very important in studying the evolution of HAeBe stars as it gives an idea about the evolution of the CSM. The equation defining the spectral index \citep{LAda1987, Wilking1989, Green1994} is  expressed as, 

\begin{equation}
n_{\lambda_1-\lambda_2} = \frac{log(\frac{\lambda_2F_{\lambda_2}}{\lambda_1F_{\lambda_1}})}{log(\frac{\lambda_2}{\lambda_1})}
\end{equation}

For our analysis we consider the spectral index, $n_{2-4.6}$, which is the ratio of the flux values at 2MASS \citep{Skrutskie2006} $K_{s}$-band (i.e., $\lambda_1$ = 2.159 $\mu$m) and WISE \citep{Cutri2013} W2-band (i.e., $\lambda_2$ = 4.6 $\mu$m). The age estimates are available only for 110 stars. However, the spectral index is not calculated for the HAeBe stars CPD-61 3587B and LkHA 224 due to the unavailability of WISE magnitudes. Hence, a sample of 108 stars is used for this analysis.  

\begin{figure}[htp]
\includegraphics[width=1\columnwidth]{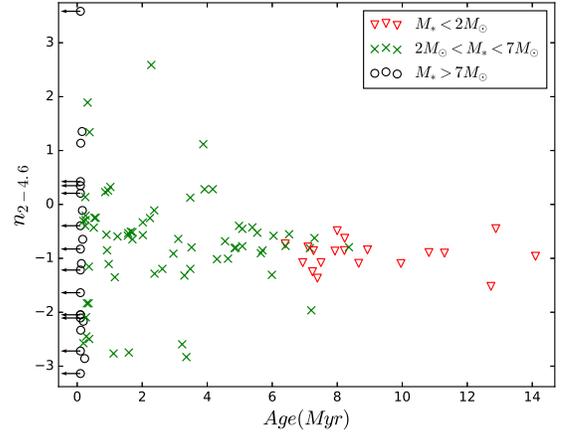}
\caption{Figure represents the graph between age and spectral index of HAeBe stars. Stars having a mass less than 2 M\textsubscript{\(\odot\)} are represented by red triangles. Stars having a mass between 2 and 7 M\textsubscript{\(\odot\)} are represented by green crosses, while stars having a mass more than 7 M\textsubscript{\(\odot\)} are shown as open circles. Arrows are assigned to the stars having an age upper limit of 0.1 Myr.}
\label{fig:6}
\end{figure}

A plot between spectral index ($n_{2-4.6}$) and age of HAeBe stars is shown in \autoref{fig:6}. No clear trend is evident in the variation of $n_{2-4.6}$ with respect to age in \autoref{fig:6}. However, when we categorize the HAeBe stars in various mass bins, a tentative trend seems to emerge. For HAeBe stars with mass less than 2 $M_{\odot}$, the $n_{2-4.6}$ value is around -1. For stars in the mass range 2$-$7 $M_{\odot}$, there is a scatter in the distribution of $n_{2-4.6}$ values, with majority of the data points around $n_{2-4.6}$ = -1. The majority of massive stars (mass $>$ 7 $M_{*}$) are showing IR index from 0.5 to -3, where the negative index is more prominent in these high mass candidates. This agrees with the study of \cite{Alonso2009} where they suggested that in high mass HBe stars disk dispersal is faster and disk masses are 5$-$10 times lesser than low mass counterparts. They explained this observation by suggesting that photoevaporation mechanism due to the UV radiation disperses the gas content in the disk, after which only a thin dusty disk containing large grains remain. The caveat in our study is the upper bound in age quoted for massive HBe stars.

\subsection{Comparison with \cite{Vioque2018}}

Calculation of stellar parameters from the theoretical HR diagram involves the use of derived variables such as bolometric luminosity (L\textsubscript{bol}) and effective temperature (T$_{eff}$). The estimation of these quantities from magnitude and color/spectral type involves approximations and comparison with standard calibration tables, which add more errors into the calculation of age and mass. Our analysis is based on the {\it Gaia} CMD rather than a theoretical HR diagram. Using a uniform photometric system combined with precise distances can give accurate estimation of age and mass of PMS stars. Thus, combining the refined stellar distances and the most consistent photometric measurements from the {\it Gaia} DR2, along with the help of synthetic photometry isochrones and evolutionary tracks from the MIST, accurate stellar ages and masses are estimated in this work. In comparison, \cite{Vioque2018} adopted the theoretical HR diagram for the analysis of age and mass. The differences between our analysis with that of \cite{Vioque2018} are listed below. 

\begin{itemize}
\item We used the photometry and distances from the {\it Gaia} for the estimation of age and mass of HAeBe stars. \cite{Vioque2018} used only the {\it Gaia} distances for the same.
\item \cite{Vioque2018} used the distance estimation method outlined in \cite{bailerjohns2018} and the calculated distances have high error bars than the values listed in the catalogue released by \cite{bailerjohns2018}. We used the distances listed in the catalog of \cite{bailerjohns2018}. For example, the distance of star DG Cir from \cite{bailerjohns2018} is $821^{+30}_{-28}$ pc. For the same star \cite{Vioque2018} estimated a distance of $833^{+52}_{-43}$ pc. 
\item We used R\textsubscript{V} = 5 for the A\textsubscript{V} calculation of HAeBe stars whereas \cite{Vioque2018} used R\textsubscript{V} = 3.1. This is because \cite{2004AJ....127.1682H} showed that total to selective extinction R\textsubscript{V} = 5 better reproduces the stellar parameters of HAeBe stars. Also, it is understood that the photometric variability and high value of reddening in HAeBe stars are not due to the interstellar medium, but due to dust particles with large grain size in the CSM \citep[see][]{Gorti1993,manoj2006}.

\item For a statistical comparison of stellar parameters with \cite{Vioque2018}, we also estimated ages and masses of HAeBe stars with R\textsubscript{V} = 3.1. The median of the fractional difference between our ages with R\textsubscript{V} = 3.1 and \cite{Vioque2018} ages is calculated to be within 19\%. The fractional difference is defined as,
\begin{equation*}
    \bigg|\frac{Vioque~estimate~-~Our~estimate}{Our~estimate}\bigg|\times100
\end{equation*}
For masses, the fractional difference is found to be within 8\%. The difference in age and mass could be due to our use of the {\it Gaia} CMD and the MIST models whereas \cite{Vioque2018} used the HR diagram and the PARSEC models \citep{ParsecBressan2012}. This comparison is extended to our actual estimates of age and mass for R\textsubscript{V} = 5. The median of the fractional difference of age and mass between our work (R\textsubscript{V} = 5) and \cite{Vioque2018} is within 31\% and 17\% respectively.

\item \cite{Vioque2018} used the H$\alpha$ EW for correlation studies with age and mass of HAeBe stars. However, for our analysis, we used the H$\alpha$ line flux, from which the $\dot{M}_{acc}$ is calculated, which is used for the correlation analysis with age and mass of HAeBe stars. It may be noted that \cite{Mendi2012} have reported that the H$\alpha$ EW may not give a clear idea about the gas content of the disk. They suggested estimating $\dot{M}_{acc}$ from the H$\alpha$ line flux to study the gas content of the disk, which we employed in this work.

\item \cite{Vioque2018} used the continuum flux distribution from 1.24 $\mu m$ to 22 $\mu m$ for the analysis of IR excess in HAeBe stars. This includes the flux measurement from the WISE W4 photometric band, which is not very reliable as the images of many HAeBe stars are not registered in W4 band. Hence, we restricted the analysis to WISE W2 band, which provides better photometry with good SNR and is free of artifacts. 
\item \cite{Vioque2018} found that there is a break in IR excess with mass. We also arrived at a similar conclusion. However, they suggested considerably low IR excess for massive HAeBe stars whereas we see a considerable range in IR excess values in this work (see \autoref{fig:6}).    

\end{itemize}

\section{Summary}

The present study made use of the unprecedented capability of the {\it Gaia} mission to derive the stellar parameters such as age and mass of HAeBe stars. Using the stellar parameters and the compiled $H\alpha$ flux, the $\dot{M}_{acc}$ for the sample is estimated. Also, we investigated the capability of the IR spectral index as a better method in quantifying the IR excess. The main results of this study are summarized below. 

\begin{itemize}
\item Better accuracy of the {\it Gaia} DR2 astrometry is confirmed from the comparison of the {\it Gaia} DR2 distances with the previously estimated values from the literature. We adopted the distance values compiled in \cite{bailerjohns2018}, which are the best distance estimates to date with minimal errors, for the sample of HAeBe stars used for this study.    

\item Age and mass of 110 HAeBe stars are estimated using the {\it Gaia} CMD, with the aid of MIST isochrones and evolutionary tracks. In our knowledge, no studies were done till now which calculated the age and mass of a confirmed sample of HAeBe stars using both the photometry and distance from the {\it Gaia} mission. Since we employed {\it Gaia} CMD for estimating the age and mass of HAeBe stars, we avoided considerable errors when these quantities are estimated from theoretical HR diagram. 

\item Mass accretion rates are calculated from the $H\alpha$ line flux measurements of 106 HAeBe stars, which is the largest sample to date. Since we had used distances and the stellar masses derived from {\it Gaia} DR2 data in the calculation of $\dot{M}_{acc}$, our estimates can be more accurate than previous studies. 

\item The disk dissipation time scale derived for our sample of HAeBe stars is $1.9 \pm 0.1$ Myr, which is consistent with the previous estimate \citep{Mendi2012}. 

\item We found that mass accretion rate is related to the mass of HAeBe stars in the form of the relation, $\dot{M}_{acc}$ $\propto$ $M_{*}^{2.8 \pm 0.2}$. 

\item We calculated the spectral index ($n_{2-4.6}$) in quantifying the IR excess in HAeBe stars. A correlation between the spectral index and age suggested a distinction between the disk of HAe and HBe stars. Massive HBe stars with ages $<$0.1 Myr show diverse values of the infrared spectral index, ranging from 0.5 to $-$3, with the negative index being more prominent. The possibility of photoevaporation resulting in the dissipation of gas content in the disk and thereby forming a thin disk and the formation difference between HBe and HAe stars needs to be explored from further studies.

\end{itemize}

\acknowledgments
We would like to thank the anonymous referee for providing
helpful comments and suggestions that improved the paper. This
work has made use of data from the European Space Agency
(ESA) mission Gaia (https://www.cosmos.esa.int/gaia), processed by the Gaia Data Processing and Analysis Consortium (DPAC; https://www.cosmos.esa.int/web/gaia/dpac/consortium). Funding for the DPAC has been provided by
national institutions, in particular, the institutions participating in
the Gaia Multilateral Agreement. Some of the data presented in
this paper were obtained from the Mikulski Archive for Space
Telescopes (MAST). STScI is operated by the Association of
Universities for Research in Astronomy, Inc., under NASA
contract NAS5-26555. Also, we made use of the VizieR catalog
access tool, CDS, Strasbourg, France.

\bibliography{example}

\begingroup
\setlength{\tabcolsep}{6pt} 
\renewcommand{\arraystretch}{1.3}
\begin{longtable}{cccccc}
\caption{Table consists of the stellar parameters for our sample of 131  HAeBe stars. The colums in the table include object name, distance, age (our work), age (\citealp{Vioque2018}-V18), mass (our work) and mass (V18). Our estimates of age and mass are derived using {\it Gaia} CMD.}
\label{tab:Table1}\\
\hline
Object & \begin{tabular}[c]{@{}c@{}}Distance \\ (pc)\end{tabular} & \begin{tabular}[c]{@{}c@{}}Age \\ (Myr)\end{tabular} & \begin{tabular}[c]{@{}c@{}}V18 Age\\ (Myr)\end{tabular} & \begin{tabular}[c]{@{}c@{}}Mass\\(M\textsubscript{\(\odot\)})\end{tabular} & \begin{tabular}[c]{@{}c@{}}V18 Mass\\(M\textsubscript{\(\odot\)})\end{tabular} \\ \hline
\endfirsthead
\multicolumn{6}{c}%
{{\bfseries Table \thetable\ continued from previous page}} \\
\hline
Object & \begin{tabular}[c]{@{}c@{}}Distance \\ (pc)\end{tabular} & \begin{tabular}[c]{@{}c@{}}Age \\ (Myr)\end{tabular} & \begin{tabular}[c]{@{}c@{}}V18 Age\\ (Myr)\end{tabular} & \begin{tabular}[c]{@{}c@{}}Mass\\(M\textsubscript{\(\odot\)})\end{tabular}& \begin{tabular}[c]{@{}c@{}}V18 Mass\\(M\textsubscript{\(\odot\)})\end{tabular} \\ \hline
\endhead
\hline
\endfoot
\endlastfoot
51 Oph & $123^{+5}_{-4}$ & $1.02^{+0.02}_{-0.02}$ & $1.22^{+0.29}_{-0.57}$ & $4^{+0.03}_{-0.02}$ & $3.35^{+0.79}_{-0.22}$ \\
AB Aur & $162^{+2}_{-2}$ & $3.92^{+0.02}_{-0.01}$ & $4^{+1.4}_{-1.5}$ & $2.34^{+0.01}_{-0.01}$ & $2.15^{+0.36}_{-0.21}$ \\
AK Sco & $140^{+1}_{-1}$ & $6.94^{+2.91}_{-2.41}$ & $8.4^{+1.7}_{-0.4}$ & $1.62^{+0.19}_{-0.17}$ & $1.401^{+0.07}_{-0.07}$ \\
AS 220 & $220^{+7}_{-7}$ & -- & $18.5^{+1.5}_{-1.4}$ & -- & $1.513^{+0.076}_{-0.076}*$ \\
AS 442 & $843^{+22}_{-21}$ & $0.26^{+0.02}_{-0.02}$ & $0.84^{+0.19}_{-0.19}$ & $6.9^{+0.22}_{-0.2}$ & $3.89^{+0.35}_{-0.26}$ \\
AS 443 & $826^{+20}_{-19}$ & $0.13^{+0.01}_{-0.01}$ & $1.13^{+0.91}_{-0.37}$ & $11.2^{+0.1}_{-0.09}$ & $3.5^{+0.48}_{-0.64}$ \\
AS 505 & $855^{+23}_{-21}$ & $0.28^{+0.01}_{-0.01}$ & $0.188^{+0.095}_{-0.065}$* & $5.8^{+0.02}_{-0.02}$ & $6.8^{+1}_{-0.9}$ \\
BD+40 4124 & $893^{+26}_{-24}$ & \textless{}0.1 & $0.1^{+0.11}_{-0.07}$ & $10.77^{+0.14}_{-0.14}$ & $9.1^{+3.9}_{-1.8}$ \\
BD+46 3471 & $759^{+17}_{-16}$ & $0.86^{+0.01}_{-0.01}$ & $1.25^{+0.64}_{-0.73}$ & $4.2^{+0.02}_{-0.01}$ & $3.3^{+1.1}_{-0.4}$ \\
BD+61 154 & $561^{+9}_{-9}$ & $0.17^{+0.01}_{-0.01}$ & $1.89^{+0.49}_{-0.78}$ & $9.5^{+0.27}_{-0.17}$ & $2.94^{+0.59}_{-0.23}$ \\
BD+65 1637 & $874^{+20}_{-19}$ & $0.3^{+0.01}_{-0.01}$ & $0.41^{+0.15}_{-0.13}$ & $6.42^{+0.02}_{-0.02}$ & $5.31^{+0.69}_{-0.48}$ \\
BF Ori & $385^{+8}_{-8}$ & $5.08^{+0.38}_{-0.36}$ & $6.38^{+0.32}_{-0.46}$ & $2.11^{+0.35}_{-0.03}$ & $1.807^{+0.09}_{-0.09}$ \\
BH Cep & $371^{+3}_{-3}$ & $12.73^{+0.21}_{-0.21}$ & $10.6^{+3}_{-3.1}$ & $1.57^{+0.01}_{-0.01}$ & $1.37^{+0.15}_{-0.1}$ \\
BO Cep & $332^{+2}_{-2}$ & $8.66^{+1.88}_{-0.43}$ & $17.1^{+0.9}_{-2.4}$ & $1.49^{+0.07}_{-0.05}$ & $1.215^{+0.061}_{-0.061}$* \\
CD-42 11721 & $1634^{+164}_{-137}$ & \textless{}0.1 & $0.023^{+0.026}_{-0.012}$* & $10.38^{+0.13}_{-0.13}$ & $20^{+7}_{-5}$ \\
CPD-61 3587B & $2672^{+303}_{-248}$ & \textless{}0.1 & -- & $13.3^{+0.2}_{-0.5}$ & -- \\
CQ Tau & $162^{+2}_{-2}$ & $10.82^{+2.49}_{-1.87}$ & $8.9^{+2.8}_{-2.5}$ & $1.65^{+0.05}_{-0.15}$ & $1.47^{+0.19}_{-0.11}$ \\
DG Cir & $821^{+30}_{-28}$ & $0.25^{+0.06}_{-0.06}$ & $4^{+16}_{-3}$ & $6.94^{+1.12}_{-0.68}$ & $2.3^{+0.6}_{-0.65}$ \\
DX Cha & $108^{+1}_{-1}$ & $3.52^{+0.02}_{-1.02}$ & $5.48^{+0.27}_{-0.4}$ & $2.48^{+0.01}_{-0.01}$ & $1.849^{+0.092}_{-0.092}$* \\
HBC 334 & $1774^{+109}_{-98}$ & $2.32^{+0.05}_{-0.05}$ & $3.71^{+0.49}_{-0.19}$ & $2.94^{+0.02}_{-0.03}$ & $2.1^{+4.3}_{-1.1}$ \\
HD 100453 & $104^{+0}_{-0}$ & $12.88^{+0.02}_{-0.03}$ & $6.53^{+0.45}_{-0.49}$ & $1.61^{+0.01}_{-0.01}$ & $1.251^{+0.063}_{-0.063}$* \\
HD 100546 & $110^{+1}_{-1}$ & $5.08^{+0.06}_{-0.08}$ & $5.5^{+1.4}_{-0.8}$ & $2.49^{+0.02}_{-0.02}$ & $2.05^{+0.1}_{-0.12}$ \\
HD 101412 & $407^{+5}_{-5}$ & $3.48^{+0.02}_{-0.02}$ & $4.37^{+0.22}_{-0.32}$ & $2.51^{+0.08}_{-0.01}$ & $2.1^{+0.11}_{-0.11}$ \\
HD 114981 & $699^{+32}_{-29}$ & $0.23^{+0.01}_{-0.01}$ & $0.277^{+0.053}_{-0.068}$* & $7.07^{+0.07}_{-0.07}$ & $6.09^{+0.59}_{-0.34}$ \\
HD 130437 & $1662^{+107}_{-95}$ & \textless{}0.1 & $0.046^{+0.077}_{-0.026}$* & $20.45^{+0.21}_{-0.2}$ & $13.4^{+4.6}_{-3.8}$ \\
HD 132947 & $378^{+8}_{-8}$ & -- & $4.05^{+0.32}_{-0.2}$ & -- & $2.22^{+0.11}_{-0.11}$ \\
HD 135344B & $135^{+1}_{-1}$ & $8.93^{+0.04}_{-0.03}$ & $8.93^{+0.45}_{-0.91}$ & $1.52^{+0.01}_{-0.01}$ & $1.432^{+0.072}_{-0.072}$* \\
HD 139614 & $134^{+1}_{-1}$ & $14.1^{+0.03}_{-0.03}$ & $14.5^{+1.4}_{-3.6}$ & $2.35^{+0.01}_{-0.01}$ & $1.481^{+0.074}_{-0.074}$* \\
HD 141569 & $110^{+1}_{-1}$ & $7.2^{+0.02}_{-0.02}$ & $9^{+11}_{-1}$ & $2.14^{+0.01}_{-0.01}$ & $1.86^{+0.093}_{-0.093}$* \\
HD 141926 & $1345^{+88}_{-78}$ & \textless{}0.1 & $0.023^{+0.007}_{-0.005}$* & \textgreater{}25 & $19.5^{+2.4}_{-2.2}$ \\
HD 142527 & $157^{+1}_{-1}$ & $2.96^{+0.02}_{-0.02}$ & $6.6^{+0.3}_{-1.5}$ & $2.4^{+0.01}_{-0.01}$ & $1.61^{+0.12}_{-0.08}$ \\
HD 142666 & $148^{+1}_{-1}$ & $7.27^{+0.08}_{-0.07}$ & $9.33^{+0.77}_{-0.47}$ & $1.82^{+0.01}_{-0.01}$ & $1.493^{+0.075}_{-0.075}$* \\
HD 144432 & $155^{+1}_{-1}$ & $7.24^{+0.02}_{-0.02}$ & $4.98^{+0.25}_{-0.55}$ & $1.81^{+0.01}_{-0.01}$ & $1.386^{+0.069}_{-0.069}$* \\
HD 145718 & $152^{+2}_{-2}$ & $5.7^{+0.17}_{-0.13}$ & $9.8^{+2.8}_{-0.5}$ & $2.09^{+0.17}_{-0.04}$ & $1.605^{+0.08}_{-0.08}$ \\
HD 150193A & $150^{+2}_{-2}$ & $4.55^{+0.03}_{-0.03}$ & $5.48^{+0.44}_{-0.27}$ & $2.2^{+0.01}_{-0.01}$ & $1.891^{+0.095}_{-0.095}$* \\
HD 163296 & $101^{+1}_{-1}$ & $6.52^{+0.26}_{-0.25}$ & $7.6^{+1.1}_{-1.2}$ & $2.1^{+0.02}_{-0.02}$ & $1.833^{+0.092}_{-0.092}$* \\
HD 179218 & $264^{+3}_{-3}$ & $2.24^{+0.01}_{-0.01}$ & $1.66^{+0.54}_{-0.26}$ & $2.95^{+0.01}_{-0.01}$ & $2.98^{+0.18}_{-0.3}$ \\
HD 190073 & $872^{+55}_{-49}$ & $0.26^{+0.01}_{-0.01}$ & $0.22^{+0.11}_{-0.07}$ & $5.99^{+0.06}_{-0.05}$ & $5.89^{+0.8}_{-0.76}$ \\
HD 200775 & $357^{+6}_{-6}$ & $0.11^{+0.01}_{-0.01}$ & $0.41^{+0.15}_{-0.2}$ & $9.41^{+0.07}_{-0.06}$ & $5.3^{+1.3}_{-0.5}$ \\
HD 216629 & $790^{+18}_{-17}$ & $0.11^{+0.01}_{-0.01}$ & $0.07^{+0.044}_{-0.033}$* & $8.22^{+0.02}_{-0.02}$ & $9.8^{+2.7}_{-1.3}$ \\
HD 244314 & $427^{+11}_{-11}$ & $7.93^{+0.05}_{-0.06}$ & $7.43^{+0.37}_{-0.54}$ & $1.77^{+0.01}_{-0.01}$ & $1.691^{+0.093}_{-0.085}$* \\
HD 244604 & $417^{+11}_{-11}$ & -- & $4.89^{+0.24}_{-0.52}$ & -- & $1.98^{+0.1}_{-0.1}$ \\
HD 245185 & $427^{+21}_{-19}$ & $5.54^{+0.22}_{-0.14}$ & $8^{+12}_{-3}$ & $2.2^{+0.01}_{-0.01}$ & $1.92^{+0.18}_{-0.1}$ \\
HD 250550 & $704^{+54}_{-47}$ & $1.7^{+0.05}_{-0.05}$ & $2.56^{+0.43}_{-0.67}$ & $3.26^{+0.04}_{-0.03}$ & $2.6^{+0.3}_{-0.14}$ \\
HD 259431 & $712^{+25}_{-23}$ & $0.25^{+0.01}_{-0.01}$ & $0.42^{+0.53}_{-0.28}$ & $6.28^{+0.04}_{-0.02}$ & $5.2^{+1.8}_{-1.3}$ \\
HD 287823 & $356^{+7}_{-7}$ & $6.04^{+0.05}_{-0.05}$ & $7.43^{+0.37}_{-0.37}$ & $2.27^{+0.01}_{-0.01}$ & $1.704^{+0.085}_{-0.085}$* \\
HD 290409 & $451^{+17}_{-16}$ & -- & $7^{+13}_{-2}$ & -- & $1.9^{+0.18}_{-0.09}$ \\
HD 290500 & $434^{+13}_{-13}$ & $8.36^{+0.17}_{-0.09}$ & $10.4^{+9.3}_{-3.3}$ & $2.04^{+0.05}_{-0.05}$ & $1.383^{+0.082}_{-0.069}$* \\
HD 290764 & $394^{+10}_{-10}$ & $6.4^{+0.06}_{-0.05}$ & $6.9^{+0.5}_{-1.4}$ & $1.88^{+0.01}_{-0.01}$ & $1.69^{+0.13}_{-0.08}$ \\
HD 290770 & $396^{+12}_{-11}$ & $4.3^{+0.11}_{-0.09}$ & $4.59^{+0.49}_{-0.54}$ & $2.39^{+0.02}_{-0.02}$ & $2.22^{+0.11}_{-0.11}$ \\
HD 305298 & $5905^{+1119}_{-829}$ & \textless{}0.1 & $0.04^{+0.31}_{-0.01}$ & $17.76^{+0.46}_{-0.48}$ & $17.7^{+2.1}_{-2}$ \\
HD 31648 & $161^{+2}_{-2}$ & $5.65^{+0.02}_{-0.02}$ & $6.2^{+0.3}_{-1.1}$ & $2.06^{+0.01}_{-0.01}$ & $1.78^{+0.13}_{-0.09}$ \\
HD 35187 & $162^{+3}_{-3}$ & $5.99^{+0.25}_{-0.94}$ & $5^{+15}_{-2}$ & $2.02^{+0.32}_{-0.12}$ & $2.1^{+0.25}_{-0.25}$ \\
HD 35929 & $384^{+8}_{-8}$ & $1.16^{+0.01}_{-0.01}$ & $1.46^{+0.07}_{-0.17}$ & $3.48^{+0.01}_{-0.01}$ & $2.92^{+0.15}_{-0.15}$ \\
HD 36112 & $160^{+2}_{-2}$ & $8^{+0.03}_{-0.04}$ & $8.3^{+0.4}_{-1.4}$ & $1.78^{+0.01}_{-0.01}$ & $1.56^{+0.11}_{-0.08}$ \\
HD 37258 & $360^{+13}_{-13}$ & $7.1^{+0.62}_{-0.03}$ & $8^{+12}_{-2}$ & $1.93^{+0.05}_{-0.1}$ & $1.88^{+0.14}_{-0.11}$ \\
HD 37357 & $796^{+297}_{-175}$ & $0.97^{+0.13}_{-0.11}$ & $1.69^{+0.87}_{-0.93}$ & $3.92^{+0.18}_{-0.16}$ & $3^{+1}_{-0.4}$ \\
HD 37490 & $320^{+45}_{-35}$ & $0.1^{+0.01}_{-0.01}$ & $0.1^{+0.11}_{-0.07}$ & $9.16^{+0.29}_{-0.23}$ & $8.6^{+3.9}_{-1.6}$ \\
HD 37806 & $423^{+11}_{-10}$ & $1.65^{+0.02}_{-0.02}$ & $1.56^{+0.64}_{-0.6}$ & $3.28^{+0.02}_{-0.02}$ & $3.11^{+0.55}_{-0.33}$ \\
HD 38120 & $402^{+14}_{-13}$ & $2.62^{+0.1}_{-0.1}$ & $3^{+14}_{-1}$ & $2.96^{+0.07}_{-0.07}$ & $2.37^{+0.43}_{-0.24}$ \\
HD 53367 & $131^{+16}_{-13}$ & -- & -- & -- & -- \\
HD 59319 & $660^{+22}_{-21}$ & $1.12^{+0.02}_{-0.02}$ & $0.96^{+0.24}_{-0.2}$ & $3.93^{+0.02}_{-0.02}$ & $3.81^{+0.31}_{-0.26}$ \\
HD 68695 & $392^{+6}_{-6}$ & $7.3^{+0.05}_{-0.06}$ & $7.6^{+1.1}_{-1.2}$ & $2.08^{+0.01}_{-0.01}$ & $1.833^{+0.092}_{-0.092}$* \\
HD 72106 & $2552^{+2141}_{-1256}$ & \textless{}0.1 & $2.1^{+2.6}_{-1.5}$ & $9.06^{+0.81}_{-0.73}$ & $2.7^{+1.5}_{-0.7}$ \\
HD 76534 & $895^{+31}_{-29}$ & $0.27^{+0.01}_{-0.01}$ & $0.171^{+0.023}_{-0.028}$* & $6.31^{+0.05}_{-0.05}$ & $7.46^{+0.51}_{-0.37}$ \\
HD 85567 & $1002^{+30}_{-28}$ & \textless{}0.1 & $0.217^{+0.045}_{-0.051}$* & $11.4^{+0.1}_{-0.1}$ & $6.32^{+0.53}_{-0.39}$ \\
HD 87403 & $2038^{+203}_{-170}$ & $0.19^{+0.01}_{-0.01}$ & $0.28^{+0.11}_{-0.08}$ & $6.72^{+0.08}_{-0.08}$ & $5.51^{+0.65}_{-0.53}$ \\
HD 94509 & $1857^{+127}_{-112}$ & -- & $0.28^{+0.17}_{-0.12}$ & -- & $5.7^{+1.1}_{-0.8}$ \\
HD 95881 & $1148^{+46}_{-42}$ & $0.16^{+0.01}_{-0.01}$ & $0.28^{+0.05}_{-0.07}$ & $7.06^{+0.04}_{-0.04}$ & $5.5^{+0.5}_{-0.27}$ \\
HD 96042 & $4007^{+649}_{-497}$ & \textless{}0.1 & $0.019^{+0.008}_{-0.005}$ & $20.09^{+0.49}_{-0.46}$ & $20.7^{+3.9}_{-2.9}$ \\
HD 97048 & $184^{+1}_{-1}$ & $3.48^{+0.01}_{-0.02}$ & $4.4^{+1.1}_{-0.3}$ & $2.52^{+0.01}_{-0.01}$ & $2.25^{+0.11}_{-0.13}$ \\
HD 98922 & $678^{+16}_{-15}$ & $0.15^{+0.01}_{-0.01}$ & $0.204^{+0.01}_{-0.038}$* & $7.42^{+0.02}_{-0.04}$ & $6.17^{+0.37}_{-0.31}$ \\
Hen 3-1191 & $1959^{+327}_{-247}$ & $0.38^{+0.03}_{-0.02}$ & $0.23^{+0.37}_{-0.11}$ & $4.96^{+0.1}_{-0.1}$ & $8.1^{+2.1}_{-0.4}$ \\
IP Per & $305^{+8}_{-7}$ & $11.3^{+0.24}_{-0.25}$ & $12^{+8}_{-3.3}$ & $1.74^{+0.01}_{-0.01}$ & $1.56^{+0.11}_{-0.12}$ \\
LkHA 167 & $1176^{+141}_{-114}$ & $0.19^{+0.01}_{-0.01}$ & $18.5^{+1.5}_{-1.4}$ & $6.18^{+0.1}_{-0.11}$ & $1.513^{+0.076}_{-0.076}$* \\
LkHA 208 & $676^{+115}_{-86}$ & $4.17^{+0.23}_{-0.22}$ & $9^{+11}_{-5}$ & $2.28^{+0.04}_{-0.05}$ & $1.56^{+0.47}_{-0.14}$ \\
LkHA 218 & $1104^{+46}_{-43}$ & $2.02^{+0.04}_{-0.04}$ & $5^{+15}_{-1}$ & $3.01^{+0.02}_{-0.02}$ & $2.12^{+0.19}_{-0.12}$ \\
LkHA 220 & $1162^{+56}_{-51}$ & $1.57^{+0.05}_{-0.05}$ & $2.04^{+0.34}_{-0.15}$ & $3.54^{+0.06}_{-0.05}$ & $3.02^{+0.15}_{-0.15}$ \\
LkHA 224 & $1253^{+249}_{-180}$ & $0.29^{+0.07}_{-0.06}$ & $1.2^{+1.1}_{-0.6}$ & $5.52^{+0.28}_{-0.12}$ & $2.85^{+0.72}_{-0.55}$ \\
LkHA 234 & $901^{+19}_{-18}$ & $0.32^{+0.04}_{-0.04}$ & $1.63^{+0.75}_{-0.6}$ & $6.38^{+0.38}_{-0.3}$ & $3.18^{+0.51}_{-0.39}$ \\
LkHA 25 & $868^{+112}_{-89}$ & -- & $6^{+14}_{-1}$ & -- & $2.3^{+0.13}_{-0.11}$ \\
LkHA 257 & $777^{+10}_{-10}$ & $7.39^{+0.03}_{-0.03}$ & $3.6^{+1.1}_{-1.1}$ & $1.82^{+0.01}_{-0.01}$ & $3.08^{+0.15}_{-0.15}$ \\
LkHA 259 & $743^{+19}_{-18}$ & $1.24^{+0.02}_{-0.02}$ & $6.4^{+1.6}_{-0.9}$ & $3.43^{+0.02}_{-0.02}$ & $1.7^{+0.1}_{-0.13}$ \\
LkHa 339 & $839^{+19}_{-18}$ & $2.37^{+0.03}_{-0.04}$ & $2.54^{+0.23}_{-0.16}$ & $3^{+0.03}_{-0.03}$ & $2.59^{+0.13}_{-0.13}$ \\
MWC 1080 & $1336^{+199}_{-154}$ & -- & $0.04^{+0.45}_{-0.02}$ & -- & $16.1^{+6.3}_{-4.2}$ \\
MWC297 & $372^{+12}_{-12}$ & -- & $0.027^{+0.006}_{-0.006}$* & -- & $16.9^{+1.9}_{-1.2}$ \\
PDS 124 & $843^{+36}_{-33}$ & $4.98^{+0.06}_{-0.07}$ & $6^{+14}_{-1}$ & $2.16^{+0.01}_{-0.01}$ & $2.07^{+0.1}_{-0.12}$ \\
PDS 130 & $1278^{+34}_{-33}$ & $2.02^{+0.02}_{--0.04}$ & $3.48^{+0.27}_{-0.26}$ & $3.12^{+0.01}_{-0.01}$ & $2.33^{+0.12}_{-0.12}$ \\
PDS 133 & $1437^{+51}_{-48}$ & $4.88^{+0.4}_{-0.43}$ & $3^{+14}_{-1}$ & $2.01^{+0.06}_{-0.04}$ & $2.93^{+0.45}_{-0.44}$ \\
PDS 134 & $2802^{+291}_{-242}$ & $0.37^{+0.02}_{-0.02}$ & $0.73^{+0.22}_{-0.21}$ & $5.62^{+0.04}_{-0.04}$ & $4.28^{+0.52}_{-0.38}$ \\
PDS 144S & $149^{+3}_{-3}$ & -- & -- & -- & -- \\
PDS 174 & $393^{+6}_{-6}$ & $3.11^{+0.04}_{-0.05}$ & $2^{+18}_{-1}$ & $2.69^{+0.02}_{-0.02}$ & $2.71^{+0.36}_{-0.23}$ \\
PDS 24 & $1099^{+23}_{-23}$ & $6.41^{+0.06}_{-0.14}$ & $10^{+10}_{-4}$ & $2.22^{+0.02}_{-0.02}$ & $1.95^{+0.1}_{-0.1}$ \\
PDS 241 & $5259^{+1535}_{-1057}$ & \textless{}0.1 & $0.078^{+0.036}_{-0.028}$* & $23.01^{+0.95}_{-0.94}$ & $11.1^{+2.3}_{-1.3}$ \\
PDS 27 & $3262^{+570}_{-428}$ & \textless{}0.1 & $0.042^{+0.072}_{-0.027}$* & \textgreater{}25 & $12.2^{+5.5}_{-3.4}$ \\
PDS 281 & $914^{+27}_{-25}$ & \textless{}0.1 & -- & $10.06^{+0.06}_{-0.04}$ & -- \\
PDS 286 & $1838^{+126}_{-111}$ & -- & $0.011^{+0.006}_{-0.001}$* & -- & $31.2^{+4.5}_{-5.5}$ \\
PDS 33 & $931^{+24}_{-23}$ & $7.16^{+0.16}_{-0.17}$ & $10.7^{+9.3}_{-3.9}$ & $2.04^{+0.01}_{-0.01}$ & $1.85^{+0.093}_{-0.093}$* \\
PDS 344 & $2360^{+96}_{-89}$ & $2.38^{+0.04}_{-0.03}$ & $1.8^{+8.4}_{-0.2}$ & $2.89^{+0.02}_{-0.02}$ & $3.48^{+0.17}_{-0.23}$ \\
PDS 361S & $3378^{+389}_{-318}$ & $0.19^{+0.01}_{-0.01}$ & $0.6^{+3.8}_{-0.3}$ & $7.81^{+0.11}_{-0.11}$ & $5^{+1}_{-0.7}$ \\
PDS 37 & $2260^{+342}_{-264}$ & -- & $0.06^{+0.1}_{-0.03}$ & -- & $10.9^{+4.5}_{-3}$ \\
PDS 415N & $144^{+3}_{-3}$ & -- & $13.1^{+5.4}_{-4.5}$ & -- & $1.21^{+0.16}_{-0.09}$ \\
PDS 431 & $1787^{+90}_{-82}$ & $1.59^{+0.03}_{-0.03}$ & $2.77^{+0.45}_{-0.73}$ & $3.46^{+0.03}_{-0.02}$ & $2.52^{+0.27}_{-0.15}$ \\
PDS 69 & $689^{+19}_{-18}$ & $0.95^{+0.02}_{-0.01}$ & $0.8^{+5.6}_{-0.3}$ & $3.96^{+0.02}_{-0.02}$ & $4.18^{+0.73}_{-0.51}$ \\
R CrA & $96^{+7}_{-6}$ & -- & -- & -- & -- \\
RR Tau & $763^{+28}_{-26}$ & $1.7^{+0.2}_{-0.64}$ & $1.98^{+0.4}_{-0.69}$ & $3.28^{+1.23}_{-0.14}$ & $2.82^{+0.46}_{-0.19}$ \\
SV Cep & $341^{+2}_{-2}$ & $4.6^{+0.24}_{-0.04}$ & $6^{+13}_{-1}$ & $2.48^{+0.11}_{-0.02}$ & $1.55^{+0.077}_{-0.077}$* \\
T Ori & $403^{+7}_{-7}$ & $4.64^{+0.36}_{-0.34}$ & $4.15^{+0.56}_{-0.67}$ & $2.18^{+0.06}_{-0.06}$ & $2.11^{+0.14}_{-0.11}$ \\
TY CrA & $136^{+3}_{-3}$ & $5.39^{+0.04}_{-0.05}$ & $6^{+14}_{-2}$ & $2.09^{+0.02}_{-0.01}$ & $2.06^{+0.22}_{-0.19}$ \\
TYC 8581-2002-1 & $549^{+7}_{-7}$ & $3.36^{+0.02}_{-0.04}$ & $8^{+12}_{-1}$ & $2.99^{+0.05}_{-0.05}$ & $1.88^{+0.094}_{-0.094}$* \\
TYC 8593-2802-1 & $1570^{+81}_{-74}$ & $3.23^{+0.06}_{-0.06}$ & $1.75^{+0.63}_{-0.35}$ & $2.43^{+0.02}_{-0.02}$ & $2.99^{+0.27}_{-0.31}$ \\
UX Ori & $322^{+5}_{-5}$ & $8.22^{+0.27}_{-0.26}$ & $11.4^{+8.6}_{-2.7}$ & $1.74^{+0.05}_{-0.02}$ & $1.612^{+0.091}_{-0.081}$* \\
UY Ori & $353^{+11}_{-10}$ & -- & -- & -- & -- \\
V1012 Ori & $383^{+8}_{-7}$ & -- & $8.5^{+1.1}_{-0.9}$ & -- & $1.3^{+0.065}_{-0.065}$* \\
V1028 Cen & $997^{+379}_{-218}$ & $2.28^{+0.26}_{-0.25}$ & $2.4^{+8.5}_{-1.1}$ & $2.93^{+0.13}_{-0.11}$ & $3^{+0.6}_{-0.15}$ \\
V1308 Ori & $5523^{+1730}_{-1168}$ & -- & $0.018^{+0.019}_{-0.008}$* & -- & $23^{+11}_{-7}$ \\
V1366 Ori & $309^{+5}_{-5}$ & -- & $6.5^{+2.4}_{-0.6}$ & -- & $1.45^{+0.072}_{-0.072}$* \\
V1787 Ori & $387^{+8}_{-8}$ & $0.57^{+0.02}_{-0.02}$ & $7.4^{+0.6}_{-1.1}$ & $2.04^{+0.03}_{-0.02}$ & $1.659^{+0.094}_{-0.083}$* \\
V346 Ori & $363^{+6}_{-6}$ & $8.23^{+0.17}_{-0.17}$ & $9.33^{+0.47}_{-0.47}$ & $1.89^{+0.01}_{-0.01}$ & $1.572^{+0.079}_{-0.079}$* \\
V350 Ori & $389^{+19}_{-18}$ & $9.96^{+1.29}_{-1.5}$ & $12.2^{+7.8}_{-4.7}$ & $1.8^{+0.08}_{-0.06}$ & $1.706^{+0.094}_{-0.085}$* \\
V380 Ori & $486^{+42}_{-36}$ & $0.51^{+0.02}_{-0.02}$ & $2^{+1}_{-0.8}$ & $5.04^{+0.08}_{-0.05}$ & $2.82^{+0.59}_{-0.38}$ \\
V599 Ori & $406^{+7}_{-7}$ & $0.36^{+0.05}_{-0.05}$ & $4.29^{+0.42}_{-0.54}$ & $2.17^{+0.15}_{-0.15}$ & $2.03^{+0.1}_{-0.1}$ \\
V699 Mon & $703^{+23}_{-22}$ & $0.53^{+0.02}_{-0.01}$ & $0.96^{+0.44}_{-0.3}$ & $4.99^{+0.03}_{-0.04}$ & $4^{+0.49}_{-0.48}$ \\
V791 Mon & $872^{+30}_{-28}$ & $0.92^{+0.01}_{-0.02}$ & $1^{+3.1}_{-0.3}$ & $4.28^{+0.04}_{-0.04}$ & $3.94^{+0.51}_{-0.45}$ \\
V856 Sco & $160^{+2}_{-2}$ & $3.88^{+0.66}_{-1.22}$ & -- & $2.22^{+0.24}_{-0.07}$ & -- \\
V892 Tau & $117^{+2}_{-2}$ & -- & -- & -- & -- \\
VV Ser & $415^{+8}_{-8}$ & $0.9^{+0.12}_{-0.37}$ & $2.8^{+8.1}_{-0.2}$ & $4.37^{+1.35}_{-0.36}$ & $2.89^{+0.14}_{-0.14}$ \\
VX Cas & $529^{+11}_{-10}$ & -- & $9^{+11}_{-4}$ & -- & $1.88^{+0.18}_{-0.09}$ \\
WW Vul & $497^{+9}_{-9}$ & $3.3^{+0.06}_{-0.06}$ & $5.08^{+0.84}_{-0.71}$ & $2.58^{+0.04}_{-0.03}$ & $1.95^{+0.11}_{-0.1}$ \\
XY Per & $456^{+19}_{-18}$ & $1.56^{+0.04}_{-0.05}$ & $1.95^{+0.43}_{-0.44}$ & $3.31^{+0.03}_{-0.04}$ & $2.82^{+0.29}_{-0.2}$ \\
Z CMa & $253^{+118}_{-61}$ & -- & $0.8^{+0.83}_{-0.59}$ & -- & $3.8^{+2}_{-0.8}$ \\ \hline
\end{longtable}
\raggedright (*)- The errors in our age and mass estimates are rounded off to two digits whereas those from \cite{Vioque2018} is reproduced as in their paper.

\end{document}